# Circularly polarized high harmonic beams carrying self-torque or time-dependent orbital angular momentum

Alba de las Heras,* Julio San Román, Javier Serrano, Luis Plaja, and Carlos Hernández-García

*Grupo de Investigación en Aplicaciones del Láser y Fotónica, Departamento de Física Aplicada, Universidad de Salamanca, E-37008 Salamanca, Spain*
*Unidad de Excelencia en Luz y Materia Estructuradas (LUMES), Universidad de Salamanca, Salamanca 37008, Spain*

E-mail: albadelasheras@usal.es

## Abstract

In the rapidly evolving field of structured light, the self-torque has been recently defined as an intrinsic property of light beams carrying time-dependent orbital angular momentum. In particular, extreme-ultraviolet (EUV) beams with self-torque— exhibiting a topological charge that continuously varies on the subfemtosecond timescale— are naturally produced in high-order harmonic generation (HHG) when driven by two time-delayed intense infrared vortex beams with different topological charges. Until now, the polarization state of such EUV beams carrying self-torque has been restricted to linear states due to the drastic reduction in the harmonic up-conversion efficiency with increasing the ellipticity of the driving field. In this work, we theoretically demonstrate how to control the polarization state of EUV beams carrying self-torque, from linear to circular. The extremely high sensitivity of HHG to the properties of the driving beam allows us to propose two different driving schemes to circumvent the current limitations to manipulate the polarization state of EUV beams with self-torque. Our advanced numerical simulations are complemented with the derivation of selection rules of angular momentum conservation, which enable precise tunability over the angular momentum properties of the harmonics with self-torque. The resulting high-order harmonic emission, carrying time-dependent orbital angular momentum with a custom polarization state, can expand the applications of ultrafast light-matter interactions, particularly in areas where dichroic or chiral properties are crucial, such as magnetic materials or chiral molecules.

# Introduction

Optical vortex beams are characterized by a twisting azimuthal phase distribution encoding the information of the orbital angular momentum (OAM).[1,2] The OAM of a light beam, $\hbar\ell$, is determined by the topological charge, $\ell$, which denotes the number of $2\pi$ phase shifts along the azimuthal coordinate.[3] OAM can be routinely imprinted into an infrared/visible light beam using spatial light modulators, spiral phase plates, or s-waveplates (among others). The low efficiency of these techniques in higher frequency regimes has been circumvented through the highly nonlinear process of high harmonic generation (HHG). In HHG, an intense infrared femtosecond laser pulse is focused on a gas or solid target, where high-order harmonics of the driving field are emitted.[4-6] The underlying mechanism can be understood through the semiclassical three-step model,[7,8] where first an electron is released by the laser field through tunnel ionization, then it is accelerated, and finally it recombines with the parent ion releasing higher frequency radiation. One of the most relevant aspects of HHG, is that the higher-order harmonics can be synthesized into attosecond pulses.[9-11] During the last decade, HHG has been essential in extending optical vortex beams to the extreme ultraviolet (EUV) spectral range,[12-20] where a higher temporal and spatial resolution down to the nanometer and attosecond scale can be achieved. Indeed, EUV beams carrying OAM possess the potential to extend the applications in diverse fields such as super-resolution imaging, nanoparticle manipulation, information processing, magnetic helicoidal dichroism, or the study of topological materials and nanostructures.[21-27]

The conservation of OAM rules the high harmonic up-conversion of a driving vortex beam with topological charge $\ell$, implying that the OAM carried by each harmonic order $q$ is $q\hbar\ell$.[13] This OAM scaling in HHG allows to reach very high topological charges that have been experimentally characterized using wavefront metrology,[20,28,29] interferometry,[15] or photoelectron ionization.[14] Nevertheless, the generation of high harmonic beams with a low topological charge is also possible in other configurations exploiting linear momentum conservation[15-17] or/and spin angular momentum (SAM) selection rules.[18,30] In addition, combined SAM-OAM control is paramount for the build-up of high-harmonic vector-vortex beams[31] and the tunability of the polarization state of structured high harmonic emission.[18,32-35]

A unique aspect of OAM-driven HHG is that high-order harmonics with time-dependent OAM can be naturally generated. It has been recently demonstrated that if HHG is driven by two linearly-polarized time-delayed vortex beams with different topological charges, the harmonics are generated with a continuously varying OAM. In analogy to mechanics, this property was denoted as self-torque.[36] Thus, the self-torque of light beams, $\hbar\xi$, has been revealed as an ultrafast property that can be imprinted in high-harmonic beams as a time-dependent twisted phase. It is defined in terms of the temporal derivative of the OAM:[36]

$$\hbar\xi = \hbar\frac{d\ell(t)}{dt}. \qquad (1)$$

The self-torque is manifested as a continuous temporal variation of OAM across the laser pulse, being an inherent feature of the light beam and not requiring any external agent. Remarkably, the self-torque is inherently related to an azimuthal frequency chirp, which facilitates its experimental characterization.[36] Though self-torque was initially obtained in the EUV regime through HHG,[36] other works in the literature have recently proposed the emission of light beams with time-dependent OAM in the visible/infrared regime using time-modulated metasurfaces with a linearly azimuthal frequency gradient[37] or engineering space-time coupling to achieve an azimuthal dependence on the topological charge.[38] All these works consider the generation of self-torque with linear polarization.

Early since the first HHG experiments, it was recognized that the generation of high-order harmonics with polarization different from linear was challenging due to the low efficiency of the process when driven by elliptically or circularly polarized light.[39] However, the initial difficulties in generating high harmonic beams with circular polarization are now standardly overcome by engineering a driving bicircular field in a two-color collinear scheme[40–46] or by employing a single-color noncollinear configuration.[47–49] After the first experiments yielding circularly polarized high-order harmonics with a nearly Gaussian spatial distribution,[43,44,47] the setups were optimized to achieve soft-X-ray photon energies[45,46] or isolated circularly polarized EUV attosecond pulses.[48] Later studies have also addressed the possibility of implementing the bicircular driving scheme or an extended noncollinear geometry for the generation of circularly polarized high harmonic vortex beams[18] and attosecond vortex pulses.[30]

In this work, we demonstrate that EUV self-torque beams can be generated with custom polarization through proper engineering of the driving field in HHG. We extend both the collinear bicircular driving and the noncollinear single-color scheme to generate circularly polarized high harmonic beams with time-varying OAM. Our numerical simulations demonstrate that the key element in both configurations is to structure the total driving beam with a time-dependent topological charge, which maps into a continuous attosecond temporal variation of the topological charge in each circularly polarized high-harmonic order. The intrinsic self-torque in HHG is also manifested as the slope of the azimuthal frequency chirp in the high-harmonic beam, which can be exploited for the experimental characterization of circularly polarized self-torqued high harmonic beams. By introducing the possibility of sculpting the polarization state of EUV beams carrying self-torque, we pave the way towards applications in ultrafast magnetism or chiral systems, where ultrafast time-dependent dichroism or chiral properties can be studied.

## Numerical methods to model OAM-driven HHG

The numerical simulations of HHG in an argon gas target have been computed using a propagation code based on the discrete dipole approximation. To account for the effect of transverse phase-matching,[50] we first consider randomly distributed elementary dipole emitters at the focal plane located in the generating gas medium and then apply the Maxwell propagator to obtain the high harmonic emission at the far-field EUV detectors.[51] At the atomic level, HHG is modeled in the quantum framework of the extended strong-field approximation.[52] This approach has already been benchmarked in several experiments of OAM-driven HHG.[18,19,30,31,36]

The spatial distribution of each driving beam is modeled in the paraxial approximation as a Laguerre-Gaussian mode, assuming the propagation along the $z$ axis:

$$LG_{\ell,m}(\rho,\phi,z) = E_0 \frac{W_0}{W(z)} \left(\frac{\rho}{W(z)}\right)^{|\ell|} L_m^{|\ell|} \left[\frac{2\rho^2}{W^2(z)}\right] \\ \times \exp\left(-\frac{\rho^2}{W^2(z)}\right) \exp\left(-ik\frac{\rho^2}{2R(z)} + ig(z) - i\ell\phi\right) \exp\left(-ikz\right). \qquad (2)$$

$E_0$ is the amplitude coefficient, $L_m^{|\ell|}$ the generalized Laguerre polynomial, $W_0$ the beam waist, $W(z) = W_0\sqrt{1 + (z/z_R)^2}$ the beam width, $z_R = \pi W_0^2/\lambda_0$ the Rayleigh length, $R(z) = z\left(1 + (z_R/z)^2\right)$ the phase-front radius, $g(z) = (2m + |\ell| + 1)\arctan(z/z_R)$ the Gouy phase, $k = 2\pi/\lambda_0$ the wavenumber, and $\lambda_0$ the central wavelength. $\rho$, $\phi$ and $z$ are the cylindrical spatial coordinates. The index $\ell$ determines the topological charge of the light beam. The other index, $m$, establishes the number of radial nodes and it is set to $m = 0$.

We have performed HHG simulations in two separate configurations where circularly polarized EUV beams carrying self-torque can be obtained. First, we have implemented two counter-rotating driving infrared *LG* beams in a noncollinear configuration. Each of the beams is composed of two time-delayed pulses with different OAM contributions. In the next section, we carefully describe the set-up and the parameters employed. Second, we have implemented a collinear bicircular scheme, where two bicircular pulses with different OAM content are time-delayed. Details are given in the corresponding section.

# Noncollinear scheme for the generation of circularly polarized high harmonic beams carrying self-torque

Noncollinear driving schemes have been experimentally implemented in HHG with linearly polarized vortex beams,[15–17] and circularly polarized vortex drivers with opposite OAM and SAM.[30] However, the noncollinear configuration of counter-rotating circularly polarized vortex drivers with the same topological charge ($\ell_{RCP} = \ell_{LCP} = \ell$) remains to be investigated. By considering the conservation laws of linear momentum ($\mathbf{k_q} = n_{RCP}\mathbf{k_{RCP}} + n_{LCP}\mathbf{k_{LCP}}$), SAM ($n_{RCP} - n_{LCP} = \pm 1$), OAM ($\ell_q = n_{RCP}\ell_{RCP} + n_{LCP}\ell_{LCP}$), and photon energy ($q = n_{RCP} + n_{LCP}$) for the right and left circularly polarized (LCP and RCP) noncollinear vortex drivers, the $q^{th}$ harmonic order is constituted by two counter-rotating circularly polarized spatial modes with equal topological charge $\ell_{q,RCP} = \ell_{q,LCP} = q\ell$. $n_{RCP}$ and $n_{LCP}$ denote the combination of photons from the RCP and LCP drivers.

In resemblance to the noncollinear scheme of HHG driven by counter-rotating circularly polarized Gaussian beams, the total electric field at the generation plane is linearly polarized, with a polarization pattern presenting a linear variation of the tilt angle along the dimension transverse to propagation. Therefore, the single-atom HHG picture being driven by a linearly polarized pulse remains efficient. The subsequent separation into the circular components at the far field is a consequence of the favorable linear momentum conservation, which in the case of a symmetric noncollinear geometry gives rise to two separated LCP and RCP high harmonic components centered at divergence angles $\pm\theta_q = \pm\arctan(q^{-1}\tan\theta_c)$.[47]

Figure 1 depicts the scheme of the noncollinear configuration to generate circularly polarized high-harmonic beams with self-torque. To achieve a time-dependent azimuthal phase distribution at the gas target, we set two time-delayed driving vortex pulses with increasing/decreasing topological charges in each of the noncollinear beams. In our example case, the topological charges of the incoming circularly polarized components are first $\ell_{RCP,1} = \ell_{LCP,1} = \ell_1 = 1$ and then $\ell_{RCP,2} = \ell_{LCP,2} = \ell_2 = 2$, leading to a temporal variation of OAM from $\hbar$ to $2\hbar$ in both the RCP and LCP components.

In our numerical simulations, we have considered laser pulses of 15.36 fs full-width half-maximum (FWHM) duration, a central wavelength of 800 nm, and a peak intensity of $4.9 \times 10^{13}$ W/cm$^2$ in each of the pulses (resulting in a total peak intensity of $2.0 \times 10^{14}$ W/cm$^2$). The time delay between the two pulses in each arm is set equal to the FWHM pulse duration, $t_d = 15.36$ fs. For the front driving modes $LG_{1,0}$, the beam waist is $W_1 = 140\,\mu$m, whereas the rear driving vortices, $LG_{2,0}$, present a waist of $W_2 = 99\,\mu$m to match the beam radii of the previous pulses. The half-crossing angle is set at $\theta_c = 3.44^o$.

The time-integrated intensity profile of the resulting driving beam at the generation plane is shown in Fig. 1b. The half-moon intensity shape results from the superposition of the two vortex beams within each arm. In order to obtain linear polarization at the target plane— and thus maximize the conversion efficiency—, it is paramount that the azimuthal position of the half-moon shape is the same for both RCP and LCP driving beams. The azimuthal position of each moon shape can be understood from the superposition of the two OAM modes. In particular, by considering a superposition of two Laguerre-Gaussian beams with a certain carrier-envelope phase (CEP), $\Phi_1$ and $\Phi_2$, and the same polarization and frequency

$$\begin{aligned}LG_{\ell_1,0}e^{-i\Phi_1} + LG_{\ell_2,0}e^{-i\Phi_2} = \\ = LG_{\ell_1,0}(A_1 + A_2 e^{-i(\ell_2-\ell_1)\phi + i(|\ell_2|-|\ell_1|)\arctan(z/z_R) - i(\Phi_2-\Phi_1)}),\end{aligned} \quad (3)$$

the condition of destructive interference, $\phi_{min}$ (i.e. the phase difference equals to $\pi + 2\pi n$,

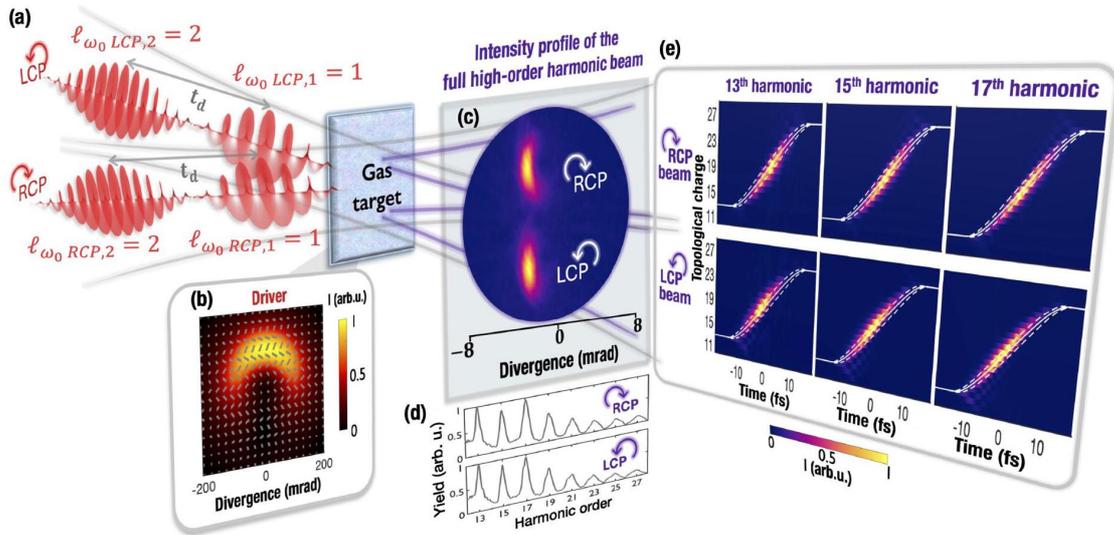

Figure 1: **Self-torqued circularly polarized high harmonic beams from a single-color noncollinear driving configuration**. (a) Scheme of the noncollinear geometry. Two counter-rotating circularly polarized beams composed of two time-delayed vortex pulses of charges $\ell_{RCP,1} = \ell_{LCP,1} = \ell_1 = 1$ and $\ell_{RCP,2} = \ell_{LCP,2} = \ell_2 = 2$ are crossed at the gas target. (b) Driving intensity profile (colormap) and polarization pattern (grey lines) at the gas target. (c) Intensity profile of the full high-harmonic beam at the EUV detectors (we integrate all the frequencies above the $11^{th}$ harmonic). (d) Spatially integrated high-harmonic spectrum of the RCP and LCP beams. (e) Temporal evolution of the topological charge of the $13^{th}$, $15^{th}$ and $17^{th}$ harmonic beams. The white lines indicate the average temporal variation of OAM and its distribution width in the analytical model.

being $n$ any integer) is satisfied at the following azimuthal positions:

$$\phi_{min} = \frac{\pi + 2\pi n + \Phi_1 - \Phi_2 + (|\ell_1| - |\ell_2|)\arctan(z/z_R)}{\ell_2 - \ell_1}. \tag{4}$$

Note that the Gouy phase term, $\arctan(z/z_R)$, introduces a phase shift of $\pi/2$ from the beam waist ($\arctan(z \to 0/z_R) = 0$) to the far-field ($\arctan(z \to \infty/z_R) = \pi/2$). For the case of $\ell_1 = 1$ and $\ell_2 = 2$ shown in Fig. 1, there is one interference position in the driving beam, and it rotates $90^o$ from the generation plane to the far-field detectors. Fig. 1b shows the intensity (colormap in the background) and polarization distribution (grey lines) when $\phi_{min}$ is synchronized for the two RCP and LCP noncollinear beams.

The half-moon profile at the peak of the interaction is also imprinted into the high-order harmonics and rotates during the harmonic beam propagation due to the Gouy phase term. The full high-order harmonic beam (constituted by the frequencies above the $11^{th}$ harmonic order) at the far-field EUV detectors is shown in Fig. 1c. We recognize two spatially separated half-moon-shaped profiles corresponding to the RCP and LCP components at the far field. The spatially integrated high-harmonic spectra for each RCP and LCP beam are represented in Fig. 1d.

In order to corroborate the generation of circularly polarized harmonics with self-torque, Fig. 1e shows the temporal variation of the topological charge for the RCP (top) and LCP (bottom) components of the $13^{th}$, $15^{th}$ and $17^{th}$ harmonic orders. It can be observed that the OAM in each harmonic order varies continuously from $q\hbar$ to $q2\hbar$ during the interaction. To obtain the temporal variation of OAM from the simulated RCP and LCP beams, we select the harmonic pulse—in a frequency range from $(q-1)\omega_0$ to $(q+1)\omega_0$—and perform the spatial Fourier transform along the azimuthal coordinate taking as the origin the center of the beam.[53] The time-dependent average OAM in the high-harmonic RCP and LCP components follows the standard law of the linearly polarized self-torque[36]

$$\hbar\overline{\ell}_{q,RCP}(t) = \hbar\overline{\ell}_{q,LCP}(t) = \hbar\overline{\ell}_q(t) = \hbar q\left[(1-\overline{\eta}(t))\ell_1 + \overline{\eta}(t)\ell_2\right], \quad (5)$$

where $\overline{\eta}(t)$ is the average of $\eta(t) = \text{Env}_2(t)/(\text{Env}_1(t) + \text{Env}_2(t))$ during the electron excursion in the HHG process. $\text{Env}_1(t)$ and $\text{Env}_2(t)$ are the envelopes of the driving pulses in each noncollinear beam. We can approximate the electron excursion in HHG to a half-cycle of the driving laser field, since the contribution from short trajectories is dominant. In this case, the width of the topological charge distribution is estimated as:[36]

$$\sigma_{\ell_q} = |\ell_2 - \ell_1|\sqrt{p_{exp}\,\overline{\eta}(t)(1-\overline{\eta}(t))}. \quad (6)$$

We set the nonperturbative exponential parameter in the harmonic amplitude to $p_{exp} = 4$, following the procedure as in ref.[36] The calculations from Eqs. (5) and (6) are superim-posed as white lines Fig. 1e. This analytical method reproduces the main features of the temporally-varying topological charge distribution.

The average self-torque of each harmonic order is defined as:[36]

$$\hbar\overline{\xi}_q = \hbar\frac{d\overline{\ell}_q(t)}{dt}, \tag{7}$$

and introduces an azimuthal frequency chirp that can be obtained by calculating the instantaneous frequency within each harmonic order[36]

$$\omega_q(t,\phi) = \frac{d\Phi_q(t,\phi)}{dt} = q\omega_0 + \frac{d\ell_q(t)}{dt}\phi = q\omega_0 + \xi_q\phi. \tag{8}$$

The results of the azimuthal frequency chirp from the simulations (colormap) and from Eq. (8) (grey dashed lines) are shown in Fig. 2. As expected, the behavior is analogous both in the LCP (Fig. 2a) and RCP (Fig. 2b) components. The value of the self-torque in Eq. (8) is approximated as:

$$\hbar\xi_q \approx \hbar\overline{\xi}_q \approx \hbar q\frac{\ell_2 - \ell_1}{3/2t_d}, \tag{9}$$

providing a close agreement with the simulations. Since the self-torque increases with the harmonic order, this implies a more pronounced slope in the azimuthal chirp for higher harmonic orders. Overall, we evidence the characterization of the azimuthal frequency chirp in the RCP and LCP components as a robust indication of the self-torque in the high harmonic beams. Note that different values of self-torque for the LCP/RCP beams could be obtained by slightly tuning the time delay in each driving beam arm.[36]

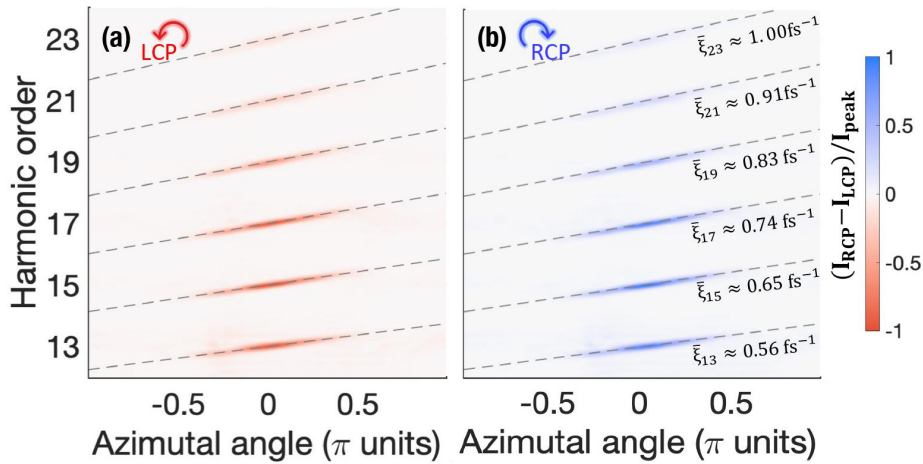

Figure 2: **Characterization of the azimuthal frequency chirp in the high-order harmonics from the single-color noncollinear configuration**. Azimuthal frequency chirp for different harmonic orders in (a) the LCP and (b) RCP noncollinear beams. The analytical calculations using Eqs. (8) and (9) are shown in grey dashed lines. The value of the average self-torque $\bar{\xi}_q$ is the same for each RCP/LCP component but different for each harmonic order, as indicated on the right.

# Collinear bicircular scheme for the generation of circularly polarized high harmonic beams carrying self-torque

A second approach to obtain circularly polarized EUV beams with self-torque relies on the implementation of a collinear bicircular driving scheme. The bicircular field is built up by superposing two counter-rotating circularly polarized beams with different frequencies. Typically, central driving frequencies of $\omega_0$ and $2\omega_0$ are considered, yielding a trefoil field structure. Bicircular fields circumvented the limitation to produce circularly polarized harmonics for the first time.[40–43] When translated into structured beams, $\omega_0$ and $2\omega_0$ LCP and RCP vortex pulse drivers of topological charges $\ell_{\omega_0}$ and $\ell_{2\omega_0}$ yield a torus-knot beam.[33,34] When driving HHG by such bicircular vortex beams, the harmonic spectrum is constituted by pairs of consecutive harmonic vortices with opposite circular polarization, followed by a suppressed harmonic order.[18,32] The topological charge of each harmonic vortex beam satisfies the following selection rule:[18]

$$\ell_q = \frac{q + 2s_q s_{\omega_0}}{3}(\ell_{\omega_0} + \ell_{2\omega_0}) - s_q s_{\omega_0}\ell_{2\omega_0}. \qquad (10)$$

The SAM of the q$^{th}$ harmonic, $s_q$, is determined by the combination of photons, $n_{\omega_0}$ and $n_{2\omega_0}$, of the driving SAM components, $s_{\omega_0}$ and $s_{2\omega_0}$, satisfying $s_{\omega_0} = -s_{2\omega_0}$ and $s_q = (n_{\omega_0} - n_{2\omega_0})s_{\omega_0}$. Note that in this scheme, the symmetry of the driving field modifies the single-atom HHG process, and thus the high-harmonic beams are generated with a different topological charge than in the noncollinear approach (where the selection rule satisfied is $\ell_{q,RCP} = \ell_{q,LCP} = q\ell$).

In order to produce circularly polarized harmonic vortices with time-dependent OAM, we consider two time-delayed torus-knot beams, whose azimuthal phase twist can be characterized with the Pancharatnam topological charge.[31,54] We set first a topological Pancharatnam charge $\ell_{P,1} = \frac{\ell_{\omega_0\,LCP,1} + \ell_{2\omega_0\,RCP,1}}{2}$, and then $\ell_{P,2} = \frac{\ell_{\omega_0\,LCP,2} + \ell_{2\omega_0\,RCP,2}}{2}$. We choose equal topological charges on the spectral components $\omega_0$ and $2\omega_0$, implying $\ell_{P,1} = \ell_{\omega_0\,LCP,1} = \ell_{2\omega_0\,RCP,1}$ and $\ell_{P,2} = \ell_{\omega_0\,LCP,2} = \ell_{2\omega_0\,RCP,2}$, so that the local field in each point of the driving beams describes the usual trefoil structure. In this configuration, the topological Pancharatnam charge of the torus-knot beam indicates the number of twists in the orientation of the trefoil along the azimuthal angle.

Figure 3a illustrates a scheme of the collinear bicircular geometry to imprint a self-torque in circularly polarized high-harmonics beams. Two counter-rotating circularly polarized beams of frequencies $\omega_0$ and $2\omega_0$ are composed of two time-delayed vortex pulses of topological charges $\ell_{2\omega_0\,RCP,1} = \ell_{\omega_0\,LCP,1} = 1$ and $\ell_{2\omega_0\,RCP,2} = \ell_{\omega_0\,LCP,2} = 2$. The driving torus-knot beams present first a topological charge $\ell_{P,1} = 1$, and then $\ell_{P,2} = 2$. In Fig. 3b, we show their intensity profile (colormap) and the electric field structure (grey lines) at the gas target. The spatial profile is analogous to a vortex beam, but the field structure describes a trefoil whose orientation changes along the azimuthal coordinate. Note that the tips of the trefoil fields describe a knotted curve embedded in the surface of a torus.[33] In the first torus-knot beam with $\ell_{P,1} = 1$, every azimuthal angle is associated with a different orientation of the trefoil. In contrast, in the second torus-knot beam with $\ell_{P,2} = 2$, the same orientation of the trefoil is repeated at azimuthal angles differing by $180^o$. The global orientation of the polarization structure is determined by the phase difference between the $\omega_0$ and $2\omega_0$ components.

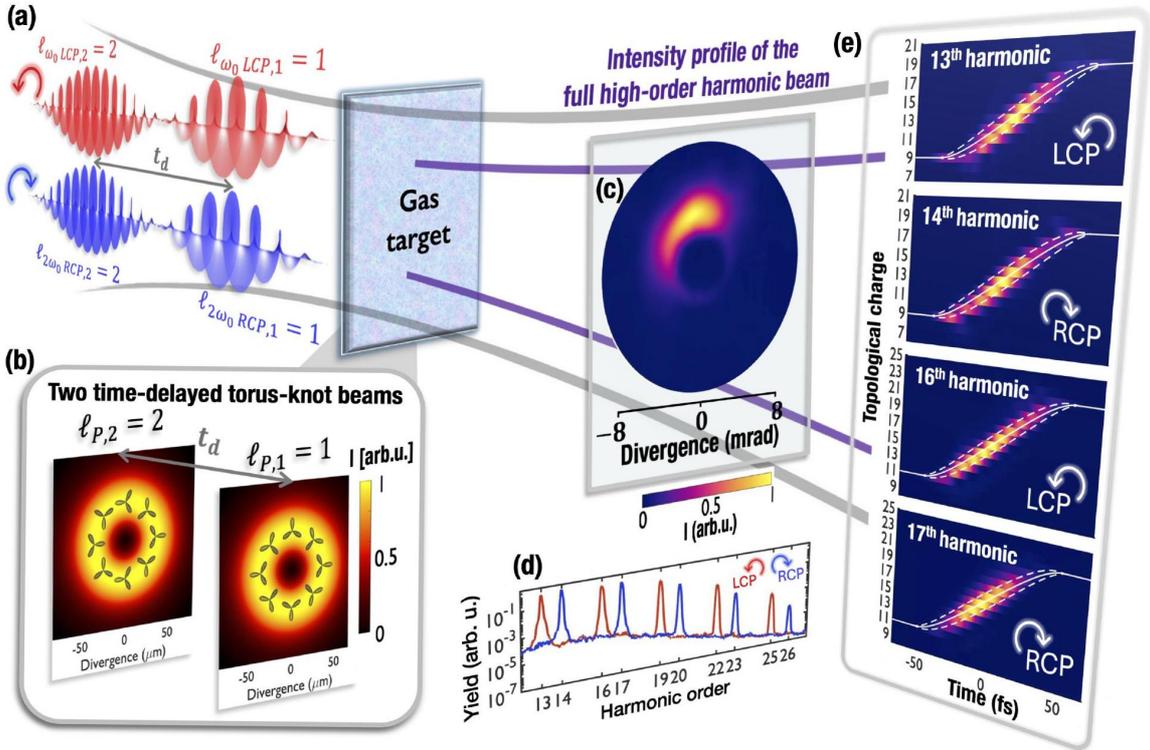

Figure 3: **Circularly polarized high harmonic beams with self-torque from a two-color collinear driving configuration**. (a) Scheme of the collinear bicircular geometry. Two counter-rotating circularly polarized beams of frequencies $\omega_0$ and $2\omega_0$ are composed of two time-delayed vortex pulses of topological charges $\ell_{2\omega_0\,RCP,1} = \ell_{\omega_0\,LCP,1} = 1$ and $\ell_{2\omega_0\,RCP,2} = \ell_{\omega_0\,LCP,2} = 2$. The total driving field is constituted by two time-delayed torus-knot beams with topological Pancharatnam charges $\ell_{P,1} = 1$ and $\ell_{P,2} = 2$. We show in (b) their intensity profile (colormap) and temporal field structure (grey lines) at the gas target. (c) Intensity profile of the full high-harmonic beam at the EUV detectors (integrating the frequencies above the $11^{th}$ harmonic). (d) Spatially integrated high-harmonic spectrum of the RCP (blue) and LCP (red) components. (e) Temporal evolution of the topological charge of the $13^{th}$, $14^{th}$, $16^{th}$ and $17^{th}$ harmonic beams. The white lines indicate the average temporal variation of OAM and its distribution width in the analytical model.

In our HHG simulations, we consider laser pulses of 57.6 fs FWHM duration, driving central wavelengths of 800 nm and 400 nm, and a peak intensity of $4.5 \times 10^{13}$ W/cm$^2$ in each of the pulses (resulting in a total peak intensity of $1.8 \times 10^{14}$ W/cm$^2$). The beam waist is $W_1 = 50.0\,\mu$m in the initial driving modes $LG_{1,0}$, and $W_2 = 35.4\,\mu$m in the final driving modes $LG_{2,0}$ to match the beam radii. Note that in this configuration, we consider longer pulse durations than in the noncollinear scheme to mimic the pulse durations usually employed in the experiments,[30,36] even if this implies a significant increase in the computational cost. The pulse duration influences the values of the self-torque but not the overall physical picture.

In our scheme, involving a time-dependent topological Pancharatnam charge in the total driving beam, the resulting polarization patterns cover more complex structures beyond the trefoil. The local field structure at each azimuthal angle is very sensitive to the synchronization between the two torus-knot beams. We set a time delay between the torus-knot beams that is equal to the FWHM pulse duration of 57.6 fs. Importantly, the carrier-envelope phase of the $2\omega_0$ pulse is adjusted to $\Phi_{2\omega,2} = -0.75\pi$. This is equivalent to setting the time delay to an integer number of $\omega_0$ cycles. This precise tunability over the time delay or the CEP of one of the pulses is required to match the trefoil patterns of the two torus-knot beams. We shall expand on this subject later on since this synchronization is crucial to efficiently generate circularly polarized harmonics with self-torque.

As in the noncollinear case, the intensity profile of the full high-harmonic beam (Fig. 3c) presents a half-moon shape. Note that it contains all the frequencies above the 11$^{th}$ harmonic, so that both RCP and LCP components are superimposed. The integrated high-harmonic spectra of the RCP (blue) and LCP (red) components are shown in Fig. 3d. By filtering specific harmonic orders, we analyze the temporal dependence of the OAM in Fig. 3e. Noticeably, the OAM in each high harmonic order evolves from $\hbar\ell_{q,1} = \hbar\dfrac{2q-s_q}{3}$ to $\hbar\ell_{q,2} = 2\hbar\dfrac{2q-s_q}{3}$, following Eq. (10).

In order to describe the average temporal variation of the OAM in this scheme, we adapt

Eq. (10) as

$$\hbar \overline{\ell}_q(t) = \hbar \left\{ \frac{q + 2s_q s_{\omega_0}}{3} \left[ \overline{\ell}_{\omega_0}(t) + \overline{\ell}_{2\omega_0}(t) \right] - s_q s_{\omega_0} \overline{\ell}_{2\omega_0}(t) \right\}, \qquad (11)$$

where we include the average of the time-dependent driving topological charges in analogy to Eq. (5),

$$\overline{\ell}_{\omega_0}(t) = (1 - \overline{\eta}_{\omega_0}(t))\ell_{\omega_0,1} + \overline{\eta}_{\omega_0}(t)\ell_{\omega_0,2} \qquad (12)$$

$$\overline{\ell}_{2\omega_0}(t) = (1 - \overline{\eta}_{2\omega_0}(t))\ell_{2\omega_0,1} + \overline{\eta}_{2\omega_0}(t)\ell_{2\omega_0,2}. \qquad (13)$$

$\overline{\eta}_{\omega_0}(t)$ and $\overline{\eta}_{2\omega_0}(t)$ are the respective averages of $\eta_{\omega_0}(t) = \mathrm{Env}_{\omega_0,2}(t)/(\mathrm{Env}_{\omega_0,1}(t) + \mathrm{Env}_{\omega_0,2}(t))$ and $\eta_{2\omega_0}(t) = \mathrm{Env}_{2\omega_0,2}(t)/(\mathrm{Env}_{2\omega_0,1}(t) + \mathrm{Env}_{2\omega_0,2}(t))$ during the electron excursion. We approximate the time of electron excursion driven by a bicircular field[42] to a representative time of 0.4 cycles of $\omega_0$. The analytical results of the average temporal variation of OAM described by (Eq. 11) and its distribution width (Eq. 6), are in perfect agreement with our numerical simulations (see white lines in Fig. 3e).

In this configuration, the average self-torque can be estimated as

$$\hbar \overline{\xi}_q \approx \hbar \frac{\ell_{q,2} - \ell_{q,1}}{3/2 t_d}. \qquad (14)$$

We now present in Fig. 4 the harmonic azimuthal frequency chirp from the numerical simulations (colormap) and the analytical calculations (grey dashed lines). In the high harmonic spectrum, each pair of counter-rotating circularly polarized harmonics exhibits the same self-torque and, thus, equal azimuthal frequency chirp. The values of the self-torque (indicated on the right of the plot) are lower than in the noncollinear geometry due to a reduced variation of OAM imposed by the simultaneous conservation of SAM and OAM, and the longer duration of the pulses considered in the simulations. By shortening the duration of the pulses (while maintaining the time delay equal to their FWHM duration), higher values of self-torque can be achieved (see Eq. 14).

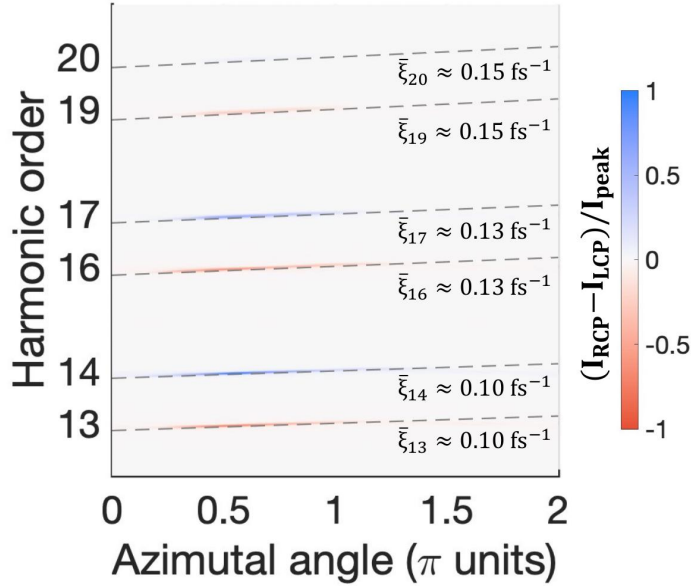

Figure 4: **Characterization of the azimuthal frequency chirp in the high-order harmonics from the two-color collinear configuration.** The analytical calculations using Eqs. (8) and (14) are shown in grey dashed lines. The value of the average self-torque $\overline{\xi}_q$, which determines the slope of the frequency chirp, is different for each harmonic order and it is indicated on the right.

Finally, it is important to note that the collinear bicircular scheme requires a precise synchronization of the driving torus-knot beams. Figure 5 analyzes two opposite cases of optimal (top row) and adverse (bottom row) overlap of the driving torus-knot beams. Panels a,d contain the intensity profile at the peak of the interaction (colormap) and the local temporal structure of the total field (grey). The second column (panels b,e) indicates the ellipticity distribution at the peak of the interaction (colormap) and the trefoil pattern of the initial (purple) and final (green) torus-knot beams. The optimal scenario corresponds to matching the trefoils of the two time-delayed torus-knot beams in the position of constructive interference, where the intensity is maximized. This occurs when the time delay is an integer number of $\omega_0$ cycles. In this case, the RCP and LCP components experience the same interference condition (Eq. 4), leading to a half-moon intensity profile in Fig. 5a and a balanced contribution of the RCP and LCP components observable in Fig. 5b. Contrarily, if the time delay is a half-integer of $\omega_0$ cycles, a ring intensity profile is preserved during the interaction and the local field describes intricate structures (see Fig 5d). We notice in Fig 5e an inhomogeneous ellipticity distribution, since the azimuthal angles of destructive interference in the LCP and RCP components differ in 180$^o$.

In Figs. 5c,f, we compare the temporally resolved OAM content of the 19$^{th}$ harmonic in both conditions, evidencing the frustration of the harmonic self-torque in the case of adverse overlap. Note that other harmonic orders experience a similar behavior. Therefore, a high-precision tunability in the time delay of less than a $\omega_0$ cycle ($\sim$ 2.67 fs) is required in an experimental setup. Alternatively, tuning the time delay within an optical cycle is equivalent to adjusting the CEP of one of the four pulses. Luckily the intensity profile of the driving beam, which can be easily retrieved in the experiments, contains the information needed to adjust the time delay or the CEP of the torus-knot beams.

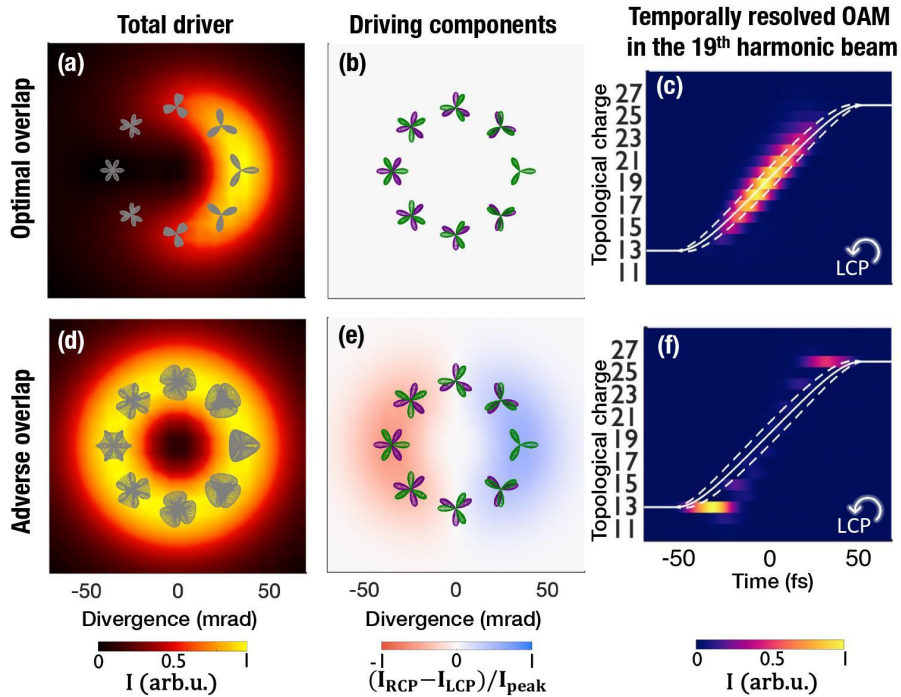

Figure 5: **Conditions of optimal (top row) and adverse overlap (bottom row) in the synchronization of the driving torus-knot beams**. In (a),(d), we represent the intensity profile at the peak of the interaction (colormap) and the local field structure (grey). Panels (b),(e) display the ellipticity distribution at the peak of the interaction (colormap) and the trefoil pattern of the initial (purple) and final (green) torus-knot beams. The synchronization of the driving torus-knot beams to a time delay that is an integer (half-integer) number of optical cycles results in an optimal (adverse) overlap of the RCP and LCP components. As shown in (c),(f) in the temporally resolved OAM content of the $19^{th}$ harmonic, the self-torque is dismantled under the condition of adverse overlap.

# Conclusion

We theoretically demonstrate the generation of high harmonic beams with circular polarization and time-dependent OAM using two different methods. First, by crossing counter-rotating circularly polarized vortex drivers at the gas target with identical topological charges, the circularly polarized harmonic components separate at the far field due to the conservation of linear momentum. In order to imprint self-torque through HHG, we create a time-dependent azimuthal phase twist in the driving beam structure. This is naturally achieved with a superposition of two time-delayed vortex beams with increasing/decreasing topological charges in each noncollinear path. The signatures of the self-torque in the noncollinear circularly polarized components of each odd high-harmonic order are a half-moon shape intensity profile, an azimuthal frequency chirp, and a continuous temporal variation of OAM along each high-harmonic pulse. This is in perfect agreement with the observations of the self-torque in linearly polarized high harmonic beams[36] since the same phenomenology is preserved.

As an alternative method, we propose a two-color bicircular collinear configuration. In this case, two driving torus-knot beams of increasing/decreasing topological Pancharatnam charges can be synchronized at the gas target to yield pairs of left and right circularly polarized harmonic orders with time-dependent OAM. The range of topological charges in each high harmonic order is different than in the noncollinear geometry, since the bicircular field symmetry maps into a different selection rule. The analytical conservation law that we derive for this configuration matches with our numerical simulations accounting for a quantum single-atom description and the macroscopic propagation of the high-harmonic beam. Even if the values of the self-torque are different than in previous configurations, the corresponding high harmonic beams are also characterized by a half-moon shape intensity profile, an azimuthal frequency chirp, and a time-dependent azimuthal phase twist associated with the continuous temporal variation of OAM along each high-harmonic pulse.

Note that in each of the two schemes proposed, we have demonstrated that circularly polarized harmonics with time-dependent OAM are obtained. A proper modification of the ellipticity of the driving beams can yield harmonic beams with self-torque and a polarization state that can range all the way from linear to circular. As relevant perspectives, we foresee that EUV circularly polarized high harmonic beams with self-torque could be applied for laser-matter interactions where dichroism or chirality is relevant. As such, these beams could be used to trigger magnetization dynamics, to perform time-resolved imaging of chiral interactions in molecules and solids, or for time-dependent quantum information processing.

In addition, these beams could be interesting for building sophisticated optical tweezers enabling a high precision control of the position and rotation of nanoparticles.

# Acknowledgements

The authors acknowledge fruitful discussions with Dr. Laura Rego, Dr. Nathan Brooks, Dr. Kevin Dorney, Prof. Margaret Murnane, Prof. Henry Kapteyn, and Prof. Thierry Ruchon. This project has received funding from the European Research Council (ERC) under the European Union's Horizon 2020 research and innovation program (Grant Agreement No. 851201). We also acknowledge the financial support from Ministerio de Ciencia de Innovación y Universidades, Agencia Estatal de Investigación and European Social Fund (PID2022-142340NB-I00). We acknowledge the computer resources at MareNostrum and the technical support provided by the Barcelona Supercomputing Center (FI-2022-3-0041, FI-2023-3-0045).

# Author contributions statement

A.H. and C. H-G. conceived the study. A.H. performed the numerical simulations and wrote the first draft of the manuscript. All authors contributed in relevant discussions about the methodology and results. All authors reviewed the manuscript and provided constructive improvements.

# References

(1) Yao, A. M.; Padgett, M. J. Orbital angular momentum: origins, behavior and applications. *Advances in Optics and Photonics* **2011**, *3*, 161.

(2) Franke-Arnold, S.; Allen, L.; Padgett, M. Advances in optical angular momentum. *Laser and Photonics Reviews* **2008**, *2*, 299–313.

(3) Allen, L.; Beijersbergen, M. W.; Spreeuw, R. J. C.; Woerdman, J. P. Orbital angular momentum of light and the transformation of Laguerre-Gaussian laser modes. *Physical Review A* **1992**, *45*, 8185–8189.

(4) McPherson, A.; Gibson, G.; Jara, H.; Johann, U.; Luk, T. S.; McIntyre, I. A.; Boyer, K.; Rhodes, C. K. Studies of multiphoton production of vacuum-ultraviolet radiation in the rare gases. *Journal of the Optical Society of America B* **1987**, *4*, 595.

(5) Ferray, M.; L'Huillier, A.; Li, X. F.; Lompre, L. A.; Mainfray, G.; Manus, C. Multiple-harmonic conversion of 1064 nm radiation in rare gases. *Journal of Physics B: Atomic, Molecular and Optical Physics* **1988**, *21*, L31–L35.

(6) Ghimire, S.; Dichiara, A. D.; Sistrunk, E.; Agostini, P.; Dimauro, L. F.; Reis, D. A. Observation of high-order harmonic generation in a bulk crystal. *Nat. Phys.* **2011**, *7*, 138–141.

(7) Schafer, K. J.; Yang, B.; DiMauro, L. F.; Kulander, K. C. Above threshold ionization beyond the high harmonic cutoff. *Physical Review Letters* **1993**, *70*, 1599–1602.

(8) Corkum, P. B. Plasma perspective on strong field multiphoton ionization. *Physical Review Letters* **1993**, *71*, 1994–1997.

(9) Farkas, G.; Tóth, C. Proposal for attosecond light pulse generation using laser induced multiple-harmonic conversion processes in rare gases. *Physics Letters A* **1992**, *168*, 447–450.

(10) Paul, P. M.; Toma, E. S.; Breger, P.; Mullot, G.; Augé, F.; Balcou, P.; Muller, H. G.; Agostini, P. Observation of a Train of Attosecond Pulses from High Harmonic Generation. *Science* **2001**, *292*, 1689–1692.

(11) Hentschel, M.; Kienberger, R.; Spielmann, C.; Reider, G. A.; Milosevic, N.; Brabec, T.;

Corkum, P.; Heinzmann, U.; Drescher, M.; Krausz, F. Attosecond metrology. *Nature* **2001**, *414*, 509–513.

(12) Zürch, M.; Kern, C.; Hansinger, P.; Dreischuh, A.; Spielmann, C. Strong-field physics with singular light beams. *Nature Physics* **2012**, *8*, 743–746.

(13) Hernández-García, C.; Picón, A.; San Román, J.; Plaja, L. Attosecond Extreme Ultraviolet Vortices from High-Order Harmonic Generation. *Physical Review Letters* **2013**, *111*, 083602.

(14) Géneaux, R.; Camper, A.; Auguste, T.; Gobert, O.; Caillat, J.; Taïeb, R.; Ruchon, T. Synthesis and characterization of attosecond light vortices in the extreme ultraviolet. *Nature Communications* **2016**, *7*, 12583.

(15) Gariepy, G.; Leach, J.; Kim, K. T.; Hammond, T. J.; Frumker, E.; Boyd, R. W.; Corkum, P. B. Creating High-Harmonic Beams with Controlled Orbital Angular Momentum. *Physical Review Letters* **2014**, *113*, 153901.

(16) Kong, F.; Zhang, C.; Bouchard, F.; Li, Z.; Brown, G. G.; Ko, D. H.; Hammond, T. J.; Arissian, L.; Boyd, R. W.; Karimi, E.; Corkum, P. B. Controlling the orbital angular momentum of high harmonic vortices. *Nature Communications* **2017**, *8*, 14970.

(17) Gauthier, D. et al. Tunable orbital angular momentum in high-harmonic generation. *Nature Communications* **2017**, *8*, 14971.

(18) Dorney, K. M.; Rego, L.; Brooks, N. J.; San Román, J.; Liao, C.-T.; Ellis, J. L.; Zusin, D.; Gentry, C.; Nguyen, Q. L.; Shaw, J. M.; Picón, A.; Plaja, L.; Kapteyn, H. C.; Murnane, M. M.; Hernández-García, C. Controlling the polarization and vortex charge of attosecond high-harmonic beams via simultaneous spin–orbit momentum conservation. *Nature Photonics* **2019**, *13*, 123–130.

(19) Pandey, A. K.; de las Heras, A.; Larrieu, T.; San Román, J.; Serrano, J.; Plaja, L.; Baynard, E.; Pittman, M.; Dovillaire, G.; Kazamias, S.; Hernández-García, C.; Guilbaud, O. Characterization of Extreme Ultraviolet Vortex Beams with a Very High Topological Charge. *ACS Photonics* **2022**, *9*, 944–951.

(20) Pandey, A. K.; de las Heras, A.; Román, J. S.; Serrano, J.; Baynard, E.; Dovillaire, G.; Pittman, M.; Durfee, C. G.; Plaja, L.; Kazamias, S.; Hernández-García, C.; Guilbaud, O. Extreme-ultraviolet structured beams via high harmonic generation. *The European Physical Journal Special Topics* **2022**, 1–10.

(21) Bliokh, K. Y. et al. Roadmap on structured waves. *Journal of Optics* **2023**, *25*, 103001.

(22) Shen, Y. et al. Roadmap on spatiotemporal light fields. *Journal of Optics* **2023**, *25*, 093001.

(23) Forbes, A.; de Oliveira, M.; Dennis, M. R. Structured light. *Nature Photonics* **2021**, *15*, 253–262.

(24) Shen, Y.; Wang, X.; Xie, Z.; Min, C.; Fu, X.; Liu, Q.; Gong, M.; Yuan, X. Optical vortices 30 years on: OAM manipulation from topological charge to multiple singularities. *Light: Science and Applications* **2019**, *8*, 90.

(25) Rubinsztein-Dunlop, H. et al. Roadmap on structured light. *Journal of Optics* **2017**, *19*, 013001.

(26) Fanciulli, M. et al. Observation of Magnetic Helicoidal Dichroism with Extreme Ultraviolet Light Vortices. *Physical Review Letters* **2022**, *128*, 077401.

(27) Shi, X.; Liao, C.-T.; Tao, Z.; Cating-Subramanian, E.; Murnane, M. M.; Hernández-García, C.; Kapteyn, H. C. Attosecond light science and its application for probing quantum materials. *Journal of Physics B: Atomic, Molecular and Optical Physics* **2020**, *53*, 184008.

(28) Sanson, F. et al. Hartmann wavefront sensor characterization of a high charge vortex beam in the extreme ultraviolet spectral range. *Optics Letters* **2018**, *43*, 2780.

(29) Sanson, F. et al. Highly multimodal structure of high topological charge extreme ultraviolet vortex beams. *Optics Letters* **2020**, *45*, 4790.

(30) de las Heras, A.; Schmidt, D.; San Román, J.; Serrano, J.; Adams, D.; Plaja, L.; Durfee, C. G.; Hernández-García, C. Attosecond vortex pulse trains, *arXiv* 2402.15225. 2024; https://arxiv.org/abs/2402.15225v1.

(31) de las Heras, A.; Pandey, A. K.; San Román, J.; Serrano, J.; Baynard, E.; Dovillaire, G.; Pittman, M.; Durfee, C. G.; Plaja, L.; Kazamias, S.; Guilbaud, O.; Hernández-García, C. Extreme-ultraviolet vector-vortex beams from high harmonic generation. *Optica* **2022**, *9*, 71–79.

(32) Paufler, W.; Böning, B.; Fritzsche, S. Tailored orbital angular momentum in high-order harmonic generation with bicircular Laguerre-Gaussian beams. *Physical Review A* **2018**, *98*, 1–5.

(33) Pisanty, E.; Rego, L.; San Román, J.; Picón, A.; Dorney, K. M.; Kapteyn, H. C.; Murnane, M. M.; Plaja, L.; Lewenstein, M.; Hernández-García, C. Conservation of Torus-knot Angular Momentum in High-order Harmonic Generation. *Physical Review Letters* **2019**, *122*, 203201.

(34) Minneker, B.; Böning, B.; Weber, A.; Fritzsche, S. Torus-knot angular momentum in twisted attosecond pulses from high-order harmonic generation. *Physical Review A* **2021**, *104*, 1–12.

(35) Luttmann, M.; Vimal, M.; Guer, M.; Hergott, J. F.; Khoury, A. Z.; Hernández-García, C.; Pisanty, E.; Ruchon, T. Nonlinear up-conversion of a polarization Möbius strip with half-integer optical angular momentum. *Science advances* **2023**, *9*, eadf3486.

(36) Rego, L.; Dorney, K. M.; Brooks, N. J.; Nguyen, Q. L.; Liao, C.-T.; San Román, J.; Couch, D. E.; Liu, A.; Pisanty, E.; Lewenstein, M.; Plaja, L.; Kapteyn, H. C.; Murnane, M. M.; Hernández-García, C. Generation of extreme-ultraviolet beams with time-varying orbital angular momentum. *Science* **2019**, *364*, eaaw9486.

(37) Sedeh, H. B.; Salary, M. M.; Mosallaei, H. Time-varying optical vortices enabled by time-modulated metasurfaces. *Nanophotonics* **2020**, *9*, 2957–2976.

(38) de Oliveira, M.; Ambrosio, A. Sub-cycle modulation of light's Orbital Angular Momentum. **2024**,

(39) Budil, K. S.; Salières, P.; Perry, M. D.; L'Huillier, A. Influence of ellipticity on harmonic generation. *Physical Review A* **1993**, *48*, 3437–3440.

(40) Eichmann, H.; Egbert, A.; Nolte, S.; Momma, C.; Wellegehausen, B.; Becker, W.; Long, S.; McIver, J. K. Polarization-dependent high-order two-color mixing. *Physical Review A* **1995**, *51*, R3414.

(41) Long, S.; Becker, W.; McIver, J. K. Model calculations of polarization-dependent two-color high-harmonic generation. *Physical Review A* **1995**, *52*, 2262.

(42) Milošević, D. B.; Becker, W.; Kopold, R. Generation of circularly polarized high-order harmonics by two-color coplanar field mixing. *Physical Review A* **2000**, *61*, 063403.

(43) Fleischer, A.; Kfir, O.; Diskin, T.; Sidorenko, P.; Cohen, O. Spin angular momentum and tunable polarization in high-harmonic generation. *Nature Photonics* **2014**, *8*, 543–549.

(44) Kfir, O.; Grychtol, P.; Turgut, E.; Knut, R.; Zusin, D.; Popmintchev, D.; Popmintchev, T.; Nembach, H.; Shaw, J. M.; Fleischer, A.; Kapteyn, H.; Murnane, M.; Cohen, O. Generation of bright phase-matched circularly-polarized extreme ultraviolet high harmonics. *Nature Photonics* **2015**, *9*, 99–105.

(45) Fan, T. et al. Bright circularly polarized soft X-ray high harmonics for X-ray magnetic circular dichroism. *Proceedings of the National Academy of Sciences* **2015**, *112*, 14206–14211.

(46) Dorney, K. M.; Fan, T.; Nguyen, Q. L. D.; Ellis, J. L.; Hickstein, D. D.; Brooks, N.; Zusin, D.; Gentry, C.; Hernández-García, C.; Kapteyn, H. C.; Murnane, M. M. Bright, single helicity, high harmonics driven by mid-infrared bicircular laser fields. *Optics Express* **2021**, *29*, 38119.

(47) Hickstein, D. D. et al. Non-collinear generation of angularly isolated circularly polarized high harmonics. *Nature Photonics* **2015**, *9*, 743–750.

(48) Huang, P.-C. et al. Polarization control of isolated high-harmonic pulses. *Nature Photonics* **2018**, *12*, 349–354.

(49) Chang, K.-Y.; Huang, L.-C.; Asaga, K.; Tsai, M.-S.; Rego, L.; Huang, P.-C.; Mashiko, H.; Oguri, K.; Hernández-García, C.; Chen, M.-C. High-order nonlinear dipole response characterized by extreme ultraviolet ellipsometry. *Optica* **2021**, *8*, 484–492.

(50) Hernández-García, C.; Sola, I. J.; Plaja, L. Signature of the transversal coherence length in high-order harmonic generation. *Physical Review A* **2013**, *88*, 43848.

(51) Hernández-García, C.; Pérez-Hernández, J. A.; Ramos, J.; Jarque, E. C.; Roso, L.; Plaja, L. High-order harmonic propagation in gases within the discrete dipole approximation. *Physical Review A* **2010**, *82*, 033432.

(52) Pérez-Hernández, J. A.; Plaja, L. Quantum description of the high-order harmonic generation in multiphoton and tunneling regimes. *Physical Review A* **2007**, *76*, 023829.

(53) Yao, E.; Franke-Arnold, S.; Courtial, J.; Barnett, S.; Padgett, M. Fourier relationship between angular position and optical orbital angular momentum. *Optics Express* **2006**, *14*, 9071.

(54) Niv, A.; Biener, G.; Kleiner, V.; Hasman, E. Manipulation of the Pancharatnam phase in vectorial vortices. *Optics Express* **2006**, *14*, 4208.